\documentstyle[11pt,appb,epsf,rotating,axodraw]{article}

\def\runhead{}

\catcode`@=11
\def\citer{\@ifnextchar [{\@tempswatrue\@citexr}{\@tempswafalse\@citexr[]}}
 
\def\@citexr[#1]#2{\if@filesw\immediate\write\@auxout{\string\citation{#2}}\fi
  \def\@citea{}\@cite{\@for\@citeb:=#2\do
    {\@citea\def\@citea{--\penalty\@m}\@ifundefined
       {b@\@citeb}{{\bf ?}\@warning
       {Citation `\@citeb' on page \thepage \space undefined}}%
\hbox{\csname b@\@citeb\endcsname}}}{#1}}
\catcode`@=12
 
\newcommand{\beq}{\begin{equation}}
\newcommand{\eeq}{\end{equation}}
\newcommand{\bea}{\begin{eqnarray}}
\newcommand{\eea}{\end{eqnarray}}

\newcommand{\tgb}{\mbox{tg$\beta$}}

\newbox\mycount
\newcommand{\ctowidth}[2]{ \setbox\mycount=\hbox{$#2$}
                          \hbox to \wd\mycount{$ \hss #1 \hss $} }
\newcommand{\ltowidth}[2]{ \setbox\mycount=\hbox{$#2$}
                          \hbox to \wd\mycount{$\hskip0pt plus0pt minus1fil
                           #1 \hfill $} }
\newcommand{\rtowidth}[2]{ \setbox\mycount=\hbox{$#2$}
                          \hbox to \wd\mycount{$\hfill #1
                          \hskip0pt plus0pt minus1fil$} }
\def\draftdate{\relax}
\def\mda{\relax}
\def\mua{\relax}
\def\mla{\relax}
\def\draft{
\def\thtystars{******************************}
\def\sixtystars{\thtystars\thtystars}
\typeout{}
\typeout{\sixtystars**}
\typeout{* Draft mode!
         For final version remove \protect\draft\space in source file *}
\typeout{\sixtystars**}
\typeout{}
\def\draftdate{\today}
\def\mua{\marginpar[\boldmath\hfil$\uparrow$]%
                   {\boldmath$\uparrow$\hfil}%
                    \typeout{marginpar: $\uparrow$}\ignorespaces}
\def\mda{\marginpar[\boldmath\hfil$\downarrow$]%
                   {\boldmath$\downarrow$\hfil}%
                    \typeout{marginpar: $\downarrow$}\ignorespaces}
\def\mla{\marginpar[\boldmath\hfil$\rightarrow$]%
                   {\boldmath$\leftarrow $\hfil}%
                    \typeout{marginpar: $\leftrightarrow$}\ignorespaces}
\def\Mua{\marginpar[\boldmath\hfil$\Uparrow$]%
                   {\boldmath$\Uparrow$\hfil}%
                    \typeout{marginpar: $\Uparrow$}\ignorespaces}
\def\Mda{\marginpar[\boldmath\hfil$\Downarrow$]%
                   {\boldmath$\Downarrow$\hfil}%
                    \typeout{marginpar: $\Downarrow$}\ignorespaces}
\def\Mla{\marginpar[\boldmath\hfil$\Rightarrow$]%
                   {\boldmath$\Leftarrow $\hfil}%
                    \typeout{marginpar: $\Leftrightarrow$}\ignorespaces}
\overfullrule 5pt
\oddsidemargin -15mm
\marginparwidth 29mm
}

\begin{document}

\renewcommand{\thefootnote}{\fnsymbol{footnote}}

\vspace*{-3cm}

\begin{flushright}
CERN-TH/98-184 \\
hep-ph/9806304 \\
June 1998
\end{flushright}

\vspace*{1cm}

\begin{center}

{\Large\sc Two-loop QCD corrections to\\[0.5cm] 
Higgs-pair production at the LHC%
\footnote{To appear in the Proceedings of the Zeuthen Workshop
on Elementary Particle Physics
{\it Loops and Legs in Gauge Theories},
Rheinsberg, 19--24 April 1998.} }

\vspace{1cm}

{\sc S.\ Dawson$^1$, S.\ Dittmaier$^2$ and M.\ Spira\footnote{Heisenberg
Fellow.}$^3$} \\
\vspace{1cm}

$^1$ {\it Physics Department\footnote{
Supported by U.S. Department of Energy contract number
DE-AC02-98CH10886.}, Brookhaven National Laboratory, \\
Upton, NY 11973, USA} \\[0.5cm]

$^2$ {\it Theory Division, CERN, CH--1211 Geneva 23, Switzerland}
\\[0.5cm]

$^3$ {\it II.\ Institut f\"ur Theoretische Physik\footnote{Supported by
Bundesministerium f\"ur Bildung und Forschung (BMBF), Bonn, Germany, under
Contract 05~7~HH~92P~(5), and by EU Program {\it Human Capital and Mobility}
through Network {\it Physics at High Energy Colliders} under Contract
CHRX--CT93--0357 (DG12 COMA).}, Universit\"at Hamburg, \\
Luruper Chaussee 149, D--22761 Hamburg, Germany}

\end{center}

\vspace{1cm}

\centerline{\bf Abstract}
\vspace{3mm}
\normalsize
\noindent
The calculation of the next-to-leading order QCD corrections to the
production of a neutral Higgs-boson pair is briefly summarized. 
The dominant production mechanism is gluon fusion via top-quark
loops, so that the QCD corrections can be considered, to good
approximation, in the limit of a heavy top-quark mass.

%\enlargethispage{2cm}

\clearpage

%\setcounter{page}{1}
%
%\title{\Large\sc Two-loop QCD corrections to\\[0.5cm] 
%	Higgs-pair production at the LHC}
%
%\author{S.\ Dawson, 
%	\address{Physics Department, Brookhaven National Laboratory, \\
%	Upton, NY 11973, USA} \\\vskip6pt
%	S.\ Dittmaier
%	\address{Theory Division, CERN, CH--1211 Geneva 23, Switzerland}
%	\and M.\ Spira
%	\address{II.\ Institut f\"ur Theoretische Physik, Universit\"at 
%	Hamburg, \\ Luruper Chaussee 149, D--22761 Hamburg, Germany} }
%
%\maketitle
% 
%\begin{abstract}
%The calculation of the next-to-leading order QCD corrections to the
%production of a neutral Higgs-boson pair is briefly summarized. 
%The dominant production mechanism is gluon fusion via top-quark
%loops, so that the QCD corrections can be considered, to good
%approximation, in the limit of a heavy top-quark mass.
%\end{abstract}

\renewcommand{\thefootnote}{\arabic{footnote}}

\setcounter{footnote}{0}

\section{Introduction}
%        ============
The search for Higgs bosons and the investigation of their properties
will be one of the most important tasks in future high-energy collider
experiments. Direct information on the Higgs potential, which reveals 
the details of the mass-generation mechanism in spontaneously broken 
gauge theories, can only be obtained by measuring
the self-interactions of the found Higgs boson(s), i.e.\ by searching
for multiple-Higgs production processes. 
In this article we briefly summarize the results of Ref.~\cite{da98}, in
which we discussed the QCD corrections to the production of neutral 
Higgs-boson pairs in $pp$ collisions at next-to-leading order (NLO)
in the framework of both the Standard Model (SM) and its minimal 
supersymmetric extension (MSSM). 

The most important production mechanism of Higgs-boson pairs in $pp$ 
collisions is gluon fusion through a heavy-quark loop, i.e.\ a NLO prediction
requires a two-loop calculation. The calculational standard for
multiloop diagrams is, however, not high enough yet to allow for an
exact analytical evaluation of massive two-loop vertex and box diagrams.
Fortunately, the structure of the relevant diagrams admits a feasible
approximation that is also phenomenologically reasonable. It consists in
an asymptotic expansion in a heavy top-quark mass, $m_t\to\infty$, and
the neglect of the $b$-quark mass, $m_b=0$. Therefore, we have calculated 
the relative NLO corrections in this limit, but taking into account the 
known leading-order (LO) cross section \cite{gghhlo} exactly. The analogous
procedure for single-Higgs production at LHC energies reproduces the 
exactly known two-loop result within 5\% for Higgs-boson masses below 
the $2m_t$ threshold of the top-quark loops; even for Higgs-boson masses
up to 1~TeV the deviation does not exceed 10\%. This impressive
reliability is due to the dominance of
universal soft-gluon effects \cite{soft}, which do not resolve the detailed 
structure of the hard-scattering process and are fully included in the
approximation. Therefore, we expect that our results for Higgs-pair
production are valid at the 10\% level, at least for invariant masses of
the Higgs-boson pair below $2m_t$.

For the production of Higgs-boson pairs in the MSSM a few more remarks
are in order. The neglect of $m_b$ forces us to choose 
$\tgb$ small, since the $b$-quark Yukawa coupling is enhanced, and thus
non-negligible, for large $\tgb$. Moreover, we consistently neglect 
squark-loop contributions, i.e.\ we assume squark masses to be large.
The production channels for a scalar and a pseudoscalar Higgs-boson
pair, $hA$ and $HA$, represent an exceptional case, because they
receive also non-negligible contributions from the Drell--Yan-like 
mechanism via $q\bar q$ fusion, which are, however, well known at LO and
NLO. More details can be found in Ref.~\cite{da98}.
\vspace{1em}

\section{Calculational framework}
The LO predictions for the gluon-fusion processes $gg\to\phi_1\phi_2$,
where $\phi_1\phi_2$ stands for any neutral Higgs-boson pair, via quark
loops are known, including the exact dependence on the heavy-quark
masses \cite{gghhlo}. The following one-loop diagrams contribute:
\\
\begin{picture}(100,80)(-30,-8)
\SetScale{0.6}
\Gluon(0,100)(50,100){3}{6}
\Gluon(0,0)(50,0){3}{6}
\ArrowLine(50,100)(100,50)
\ArrowLine(100,50)(50,0)
\ArrowLine(50,0)(50,100)
\DashLine(100,50)(150,50){5}
\DashLine(150,50)(200,100){5}
\DashLine(150,50)(200,0){5}
\put(128,57){$\phi_1$}
\put(128, 0){$\phi_2$}
\put(67,38){$\phi,Z$}
\put(-15,0){$g$}
\put(-15,57){$g$}
\put(10,28){$t,b$}
\SetScale{1}
\end{picture}
\begin{picture}(100,80)(-100,-8)
\SetScale{0.6}
\Gluon(0,100)(50,100){3}{6}
\Gluon(0,0)(50,0){3}{6}
\ArrowLine(50,100)(140,100)
\ArrowLine(140,100)(140,0)
\ArrowLine(140,0)(50,0)
\ArrowLine(50,0)(50,100)
\DashLine(140,100)(190,100){5}
\DashLine(140,0)(190,0){5}
\put(125,57){$\phi_1$}
\put(125, 0){$\phi_2$}
\put(-15, 0){$g$}
\put(-15,57){$g$}
\put(10,28){$t,b$}
\SetScale{1}
\end{picture}
\\
where the $Z$-boson exchange in the $s$-channel is only relevant for
$hA$ and $HA$ production. At NLO, we get virtual corrections formed by 
two-loop diagrams, which contain a closed quark loop and one, two or 
three internal gluons. Moreover, there are real one-loop corrections
induced by the parton processes $gg\to\phi_1\phi_2 g$,
$gq\to\phi_1\phi_2 q$, and $q\bar q\to\phi_1\phi_2 g$.
After summing virtual and real corrections, IR divergences drop out; the
remaining collinear singularities are absorbed in the parton
distributions when calculating the $pp$ cross section.

For the heavy-mass expansion of the diagrams containing the top quark we
employ the general algorithm of Ref.~\cite{sm97}. The actual expansion
is carried out in the integrand of a diagram, i.e.\ before the
momentum-space integration, and relies on dimensional
regularization. The general algorithm is easily summarized.
Given any Feynman graph $\Gamma$ with some large
internal masses $M_i$, and denoting the corresponding 
amplitude and integrand by $F_\Gamma$ and $I_\Gamma$, respectively,
the large-mass expansion reads
\def\asymp#1{\mathrel{\raisebox{-.4em}{$\widetilde{\scriptstyle #1}$}}}
\beq
\textstyle
F_\Gamma \; = \; \int \left(\prod_l d^n q_l\right) \, I_\Gamma
\;\; \asymp{M_i\to\infty} \;\;
\sum_\gamma \, \int \left(\prod_l d^n q_l\right) \,
I_{\Gamma/\gamma} \, {\cal T}_{p^\gamma_i,m_i} I_\gamma,
\label{eq:Fexpand}
\eeq
where $q_l$ are the integration momenta.
The sum on the r.h.s.\ runs over all subgraphs $\gamma$ of $\Gamma$
that contain all propagators with the heavy masses $M_i$ and that are
irreducible with respect to those lines of $\gamma$
that carry light masses
$m_i$. The integrand of the subgraph $\gamma$ is denoted by $I_\gamma$.
The reduced graph $\Gamma/\gamma$ results from $\Gamma$ upon
shrinking $\gamma$ to a point, and the integrand $I_{\Gamma/\gamma}$ is
defined such that $I_\Gamma=I_\gamma I_{\Gamma/\gamma}$. The symbol
${\cal T}_{p^\gamma_i,m_i}$ represents an operator that replaces the
integrand $I_\gamma$ by its Taylor series in the expansion parameters
$p^\gamma_i$ and $m_i$, where $p^\gamma_i$ are the external momenta of
the subgraph $\gamma$. Therefore, Eq.~(\ref{eq:Fexpand}) expresses the
original integral $F_\Gamma$ by a sum over simpler integrals with a finite
number of terms in each order $M_i^a$.

As an illustrating example, we inspect the following box diagram.
\\
\centerline{
\begin{picture}(180,75)(-20,-5)
%\put(180, 90){\rule{2mm}{2mm}}
%\put(-20, 90){\rule{2mm}{2mm}}
%\put(-20,-20){\rule{2mm}{2mm}}
%\put(180,-20){\rule{2mm}{2mm}}
\SetScale{.6}
\Gluon(0,100)(50,100){3}{6}
\Gluon(0,0)(50,0){3}{6}
\ArrowLine(95,0)(95,100)
\Gluon(50,100)( 95,100){3}{6}
\ArrowLine(95,100)(140,100)
\ArrowLine(140,100)(140,0)
\ArrowLine(140,0)(95,0)
\Gluon(50,0)( 95,0){3}{6}
\Gluon(50,0)(50,100){3}{12}
\DashLine(140,100)(190,100){5}
\DashLine(140,0)(190,0){5}
\put(125,57){$\phi_1$}
\put(125, 0){$\phi_2$}
\put(-15, 0){$g$}
\put(-15,57){$g$}
\put( 90,30){$t$}
\SetScale{1}
\end{picture} }
\\
There are two subgraphs $\gamma$ that are relevant in the
expansion (\ref{eq:Fexpand}) of this graph. The first one,
$\gamma_{\mathrm{loop}}$, consists of all internal lines, the second
one, $\gamma_{\mathrm{top}}$, consists of the top-quark lines only. The
reduced graphs look as follows.
\\
\begin{picture}(120,75)(-75,-5)
\SetScale{0.6}
\Gluon(0,100)(50,55){3}{6}
\Gluon(0,0)(50,50){3}{6}
\DashLine(50,50)(100,100){5}
\DashLine(50,50)(100,0){5}
\GCirc(50,50){12}{0.5}
\put(65,57){$\phi_1$}
\put(65, 0){$\phi_2$}
\put(-15, 0){$g$}
\put(-15,57){$g$}
\put(-65,30){$\Gamma/\gamma_{\mathrm{loop}}$:}
\SetScale{1}
\end{picture} 
\begin{picture}(100,75)(-130,-5)
\SetScale{0.6}
\Gluon(0,100)(25,75){3}{3}
\Gluon(25,75)(50,50){3}{3}
\Gluon(0,0)(25,25){3}{3}
\Gluon(25,25)(50,50){3}{3}
\GlueArc(50,50)(35.355,135,225){3}{5}
\DashLine(50,50)(100,100){5}
\DashLine(50,50)(100,0){5}
\GCirc(50,50){12}{0.5}
\put(65,57){$\phi_1$}
\put(65, 0){$\phi_2$}
\put(-15, 0){$g$}
\put(-15,57){$g$}
\put(-65,30){$\Gamma/\gamma_{\mathrm{top}}$:}
\SetScale{1}
\end{picture} 
\\
In the case of $\gamma_{\mathrm{loop}}$ the Taylor-expansion operator
replaces each propagator $P(q-p,m)$ with $m=0,m_t$ by 
\begin{equation}
\textstyle
P(q-p,m) = [(q-p)^2-m^2]^{-1} =
\sum_{l=0}^\infty (q^2-m^2)^{-1-l} (2q p-p^2)^l, 
\label{eq:Prepl}
\end{equation}
where $q$ is a linear combination of the two integration momenta $q_1$
and $q_2$, and $p$ consists of external momenta $p_i$. This replacement
leads to two-loop vacuum integrals of the form
\begin{equation}
\int d^n q_1 \, \int d^n q_2 \,
\frac{q_{1,\mu_1}\dots q_{1,\mu_R} \, q_{2,\nu_1}\dots q_{2,\nu_S}}
{(q_1^2)^{n_1} (q_2^2-m_t^2)^{n_2} [(q_1+q_2)^2-m_t^2]^{n_3}},
\end{equation}
the calculation of which is straightforward in $n$ dimensions.
For $\gamma_{\mathrm{top}}$ the Taylor expansion concerns only the
top-quark propagators, and the replacement (\ref{eq:Prepl}) 
applies for $m=m_t$ and $q=q_2$, where $q_2$ is the loop momentum
running through the top-quark loop. Note that $p$ now includes all external
momenta of the process as well as the loop momentum $q_1$ running through 
the internal gluon lines. Thus, the integration over $q_2$ yields simple
one-loop vacuum integrals
\begin{equation}
\int d^n q_2 \, \frac{q_{2,\mu_1}\dots q_{2,\mu_R}}{(q_2^2-m_t^2)^{n_1}},
\label{eq:V1tens}
\end{equation}
while the $q_1$ integration leads to one-loop integrals of the form
\begin{equation}
\int d^n q_1 \, \frac{q_{1,\mu_1}\dots q_{1,\mu_R}} {q_1^2(q_1+p_1)^2},
\qquad
\int d^n q_1 \,
\frac{q_{1,\mu_1}\dots q_{1,\mu_R}} {q_1^2(q_1+p_1)^2(q_1+p_2)^2},
\end{equation}
containing massless propagators and non-vanishing external momenta $p_1$
and $p_2$.

It is interesting to realize the close connection of the diagrammatic
expansion algorithm described above and the independent approach using
a low-energy effective theory. The effective Lagrangian results from
integrating out the heavy top quark from the full theory and quantifies the 
local interactions between light Higgs bosons and gluons at low energies
induced by heavy-quark loops. The effective interactions that are
relevant in our case are known in LO as well as in NLO and are summarized in
Ref.~\cite{da98}. The effective couplings can be derived from the gluon 
self-energy upon differentiation, except for the ones that involve an
odd number of pseudoscalar Higgs bosons. The latter are related to the 
ABJ anomaly. We have also calculated the NLO corrections to 
$gg\to\phi_1\phi_2$ in the effective-Lagrangian approach.
There is a one-to-one correspondence between the sum of all
topologically equal reduced graphs $\Gamma/\gamma$ in the diagrammatical
expansion and the corresponding effective diagram in which $\gamma$ is
identified with the effective coupling. More precisely, the effective 
couplings enter the NLO corrections in two different ways. Firstly, the 
NLO parts of the effective interactions contribute in tree-level
diagrams like $\Gamma/\gamma_{\mathrm{loop}}$. Secondly, 
the LO parts of the effective couplings contribute in one-loop diagrams 
like $\Gamma/\gamma_{\mathrm{top}}$, which are the only sources of
non-local effects.

Although we set $m_b=0$, there are contributions
from $b$-quark loops in the $hA$ and $HA$ production channels. These
contributions are associated with $Z$-boson exchange and have to be
considered separately. The virtual two-loop corrections in the $Zgg$
vertex are fully detemined by gauge invariance, because only a single
form factor contributes. The divergence of the $Zgg$ vertex is
related by a Ward identity to the coupling of a pseudoscalar field to 
two gluons, implying that the $Zgg$ vertex correction concides with
the one for single $A$ production, which gets no $b$-quark contribution
for $m_b=0$. The $b$-quark loops appearing in the real corrections do not
possess a simple form. However, the dominant effects, the ones that are
sensitive to soft gluons, are already included by taking into account
the $Zgg$ vertex only, leading again to the same correction as for the
$Agg$ vertex.

\section{Results}

The final analytical results for the NLO corrections to the process 
$pp\to\phi_1\phi_2+X$ have a rather compact form in the adopted
approximation. 
In this context, the LO cross section $\hat\sigma_{\mathrm{LO}}$ of 
the parton process $gg\to\phi_1\phi_2$ is of central importance: 
\begin{equation}
\hat \sigma_{\mathrm{LO}}(\hat s = Q^2) = \int d\hat t \,
\frac{G_F^2 \alpha_s^2(\mu)}{256 (2\pi)^3} \left\{ \left|
C_\triangle F_\triangle + C_\Box F_\Box \right|^2 + \left| C_\Box G_\Box
\right|^2 \right\},
\label{eq:gghhlo}
\end{equation}
where $Q$ is the invariant mass 
of the Higgs-boson pair, and $\hat s$, $\hat t$, $\hat u$ are the usual 
Mandelstam variables in the partonic centre-of-mass (CM) system. 
The functions $C_{\triangle,\Box}$ contain
coupling factors and $s$-channel propagators, and $F_{\triangle,\Box}$
and $G_\Box$ represent the form factors corresponding to the two tensors 
that are relevant for spin-0 and spin-2 exchange. The explicit
expressions of the $C$'s, $F$'s, and $G$'s, including the full
fermion-mass dependence for the latter two, are given in Ref.~\cite{gghhlo}.
Here we only give the $C$'s and the asymptotic form of the
form factors for the SM:
\begin{eqnarray}
C_\triangle & = & \frac{3M_H^2}{\hat s - M_H^2 + iM_H \Gamma_H}, 
\qquad
C_\Box = 1, \nonumber \\ 
F_\triangle & \to & \textstyle\frac{2}{3}, \qquad 
F_\Box \to \textstyle -\frac{2}{3}, \qquad
G_\Box \to 0 \qquad \mbox{for} \; m_t\to\infty, \; m_b=0.
\end{eqnarray}
Note, however, that the exact form of the form factors is used in the
numerical evaluations.

All contributions to the $pp$ cross section $\sigma_{\mathrm{NLO}}$,
\beq
\sigma_{\mathrm{NLO}} = \sigma_{\mathrm{LO}} + 
\Delta \sigma_{\mathrm{virt}} + \Delta \sigma_{gg} +
\Delta \sigma_{gq} + \Delta \sigma_{q\bar q},
\eeq
can be written as a convolution of a correction factor, of 
$\hat\sigma_{\mathrm{LO}}$, and of the 
parton--parton luminosities $d{\cal L}^{ij}/d\tau$:
\begin{eqnarray}
\sigma_{\mathrm{LO}} & = & \int_{\tau_0}^1 d\tau~\frac{d{\cal L}^{gg}}{d\tau}~
\hat\sigma_{\mathrm{LO}}(\tau s), 
\nonumber \\ 
\Delta \sigma_{\mathrm{virt}} & = & \frac{\alpha_s(\mu)} {\pi}\int_{\tau_0}^1 d\tau~
\frac{d{\cal L}^{gg}}{d\tau}~\hat \sigma_{\mathrm{LO}}(\tau s)~C, 
\nonumber \\ 
\Delta \sigma_{gg} & = & \frac{\alpha_{s}(\mu)} {\pi} \int_{\tau_0}^1 d\tau~
\frac{d{\cal L}^{gg}}{d\tau} \int_{\tau_0/\tau}^1 \frac{dz}{z}~
\hat\sigma_{\mathrm{LO}}(z \tau s)
\left\{ - z P_{gg} (z) \log \frac{M^{2}}{\tau s} \right. 
\nonumber \\
& & \left. \hspace*{2em} {} - \frac{11}{2} (1-z)^3 + 6 [1+z^4+(1-z)^4]
\left(\frac{\log (1-z)}{1-z} \right)_+ \right\}, 
\nonumber \\ 
\Delta \sigma_{gq} & = & \frac{\alpha_{s}(\mu)} {\pi} \int_{\tau_0}^1 d\tau
\sum_{q,\bar{q}} \frac{d{\cal L}^{gq}}{d\tau} \int_{\tau_0/\tau}^1 \frac{dz}{z}~
\hat \sigma_{\mathrm{LO}}(z \tau s)
\nonumber \\
&& \hspace*{2em}
\times \left\{ -\frac{z}{2} P_{gq}(z) \log\frac{M^{2}}{\tau s(1-z)^2} 
+ \frac{2}{3}z^2 - (1-z)^2 \right\},
\nonumber \\ 
\Delta \sigma_{q\bar q} & = & \frac{\alpha_s(\mu)}
{\pi} \int_{\tau_0}^1 d\tau
\sum_{q} \frac{d{\cal L}^{q\bar q}}{d\tau} \int_{\tau_0/\tau}^1 \frac{dz}{z}~
\hat \sigma_{\mathrm{LO}}(z \tau s)~\frac{32}{27} (1-z)^3,
\end{eqnarray}
where $\sqrt{s}$ is the total CM energy in the $pp$ system, and
$\sqrt{\tau_0 s} = m_1+m_2$ is the corresponding threshold for the 
production of two Higgs bosons with masses $m_1$ and $m_2$.
The renormalization and factorization scales are denoted by $\mu$ and
$M$, respectively, and $P_{ij}(z)$ are the Altarelli--Parisi splitting
functions.
While the form of the real corrections $\Delta \sigma_{gg,gq,q\bar q}$
is universal, the virtual correction factors 
\begin{eqnarray}
C & = & \pi^2 + c_1 + \frac{33-2N_F}{6} \log \frac{\mu^2}{Q^2} 
\\*
&& {} + \Re e~\frac{\int d\hat t 
\left\{c_2~C_\Box (C_\triangle F_\triangle + C_\Box F_\Box) 
+ \frac{\hat t\hat u-m_1^2 m_2^2}{2Q^2}
\left(\frac{c_3}{\hat t}+\frac{c_4}{\hat u}\right) 
C^2_\Box G_\Box \right\}}
{\int d\hat t \left\{ |C_\triangle F_\triangle +
C_\Box F_\Box |^2 + |C_\Box G_\Box|^2 \right\}}
%\hspace*{2em}
\nonumber 
\label{eq:cfac}
\end{eqnarray}
do not coincide for all Higgs-boson pairs. 
%The actual values of the
%numerical constants $c_i$ depend on the final state
%$\phi_1\phi_2$; and can be found in Ref.~\cite{da98}.
The coefficients $c_i$ for the individual final-state Higgs bosons
$\phi_1\phi_2$ are given by
\def\arraystretch{1.2}
\begin{eqnarray}
\begin{array}[b]{lllll}
\phi_1\phi_2=hh,hH,HH: \; & c_1 = \frac{11}{2}, \; & 
c_2 = \frac{4}{9}, \; & c_3 = \frac{4}{9}, & c_4 = \frac{4}{9}, \\ 
\phi_1\phi_2=hA,HA: & 
c_1 = 6, & c_2 = \frac{2}{3}, & c_3 = -\frac{2}{3}, & c_4 = \frac{2}{3}, \\
\phi_1\phi_2=AA: & 
c_1 = \frac{11}{2}, & c_2 = -1, & c_3 = 1, & c_4 = 1.
\end{array}
\hspace*{1em}
\end{eqnarray}

\begin{figure}[t]
\vspace*{0.0cm}
\hspace*{0.0cm}
\begin{turn}{-90}%
\epsfxsize=8.5cm \epsfbox{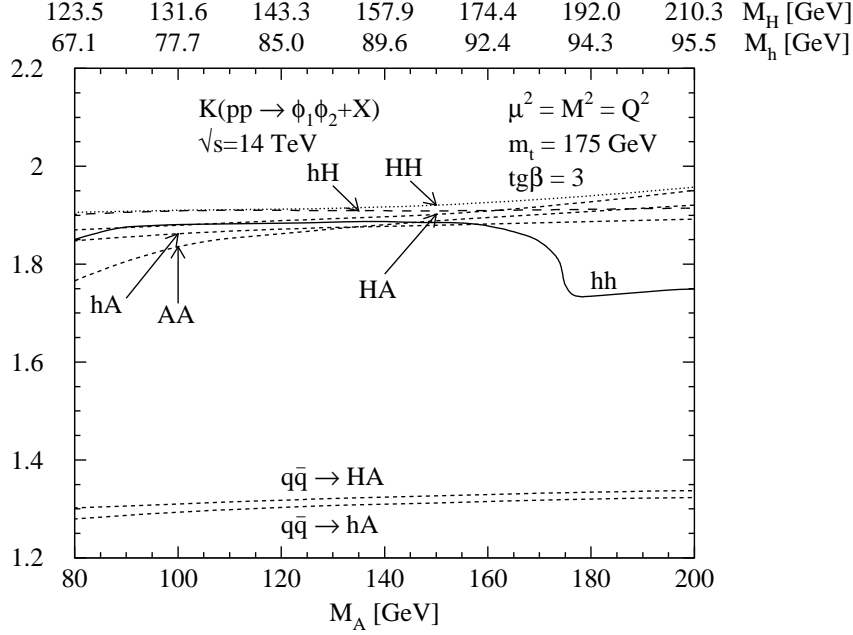}
\end{turn}
%\vspace*{-0.5cm}
\caption[]{\it \label{fg:kmssm} $K$~factors of the QCD-corrected gluon-fusion
and Drell--Yan like cross sections $\sigma(pp \to \phi_1\phi_2+X)$ at the LHC
with CM~energy $\sqrt{s}=14$~TeV. 
}
\end{figure}
A detailed numerical study of the QCD corrections described above is
presented in Ref.~\cite{da98} for the SM and the MSSM. 
Figure~\ref{fg:kmssm} summarizes our results on the $K$~factors for 
Higgs-pair production in the MSSM as functions of the pseudoscalar mass
$M_A$, demonstrating their phenomenological importance at the LHC.
The $K$~factor for the SM cross section smoothly varies between 1.9 and
2.0 for a Higgs-boson mass within 80--200~GeV.

\section*{Acknowledgement}

S.~Dittmaier would like to thank the organizers for the kind invitation
and for providing a very pleasant atmosphere during the workshop.

\end{document}